# Yttrium incorporation in $Cr_2AlC$: On the metastable phase formation and decomposition of $(Cr,Y)_2AlC$ MAX phase thin films

**Clio Azina[a*], Tim Bartsch[a], Damian M. Holzapfel[a], Martin Dahlqvist[b], Johanna Rosen[b], Lukas Löfler[a], Alba San Jose Mendez[c], Marcus Hans[a], Daniel Primetzhofer[d], Jochen M. Schneider[a]**

[a]Materials Chemistry, RWTH Aachen University, Kopernikusstraße 10, D-52074, Aachen, Germany
[b]Materials Design, Department of Physics, Chemistry and Biology (IFM), Linköping University, SE-581 83 Linköping, Sweden
[c]Deutsches Elektronen-Synchrotron DESY, Notkestr. 85, D-22607 Hamburg, Germany
[d]Department of Physics and Astronomy, Uppsala University, Lägerhyddsvägen 1, S-75120 Uppsala, Sweden

*Corresponding author; E-Mail: azina@mch.rwth-aachen.de

## Abstract

Herein we report on the synthesis of a metastable $(Cr,Y)_2AlC$ MAX phase solid solution by co-sputtering from a composite Cr-Al-C and elemental Y target, at room temperature, followed by annealing. While direct high-temperature synthesis resulted in multiphase films, as evidenced by X-ray diffraction analyses, room temperature depositions, followed by annealing to 760 °C led to the formation of phase pure $(Cr,Y)_2AlC$ by diffusion. Higher annealing temperatures caused decomposition of the metastable phase into $Cr_2AlC$, $Y_5Al_3$ , and Cr-carbides. In contrast to pure $Cr_2AlC$, the Y-containing phase crystallizes directly in the MAX phase structure instead of first forming a disordered solid solution. Furthermore, the crystallization temperature was shown to be Y-content dependent and was increased by ~200 °C for 5 at.% Y compared to $Cr_2AlC$. Calculations predicting the metastable phase formation of $(Cr,Y)_2AlC$ and its decomposition are in excellent agreement with the experimental findings.

**Keywords:** MAX phases, solid solution, crystallization, metastable phase formation, decomposition

## 1. Introduction

MAX phases constitute a promising family of inherently nanolaminated carbides and nitrides which crystallize in a hexagonal *P*6₃/*mmc* structure, and are given by the formula $M_{n+1}AX_n$, where M is a transition metal, A is an A-group element and X is either carbon and/or nitrogen (n = 1, 2, 3) [1,2]. Their structure and chemical composition provide them with a unique combination of metallic and ceramic properties [3]. MAX phases have been shown to be promising for a variety of applications, particularly extreme environments [4–9], but they are also used as precursors for the synthesis of their well-known 2D derivatives, the so-called MXenes [10,11].

The chemical versatility of MAX phases has been evidenced in the last years, by the discovery of no less than 180 different MAX phases [6]. Aside from the conventional ternary MAX phases, solid solutions have also been reported in both bulk and thin film form [12–17], on both M and A sites. These solid solutions have been contemplated to improve the oxidation resistance of ternary MAX phases, for example [18,19].

Berger *et al.* reported doping of $Cr_2AlC$ with up to 0.3 at.% of Y [20,21]. The authors reported variations of lattice parameters with increasing Y content, however, their system appeared to be composed of both $Cr_2AlC$ and the disordered hexagonal solid solution $(Cr,Al)_2C_x$. Upon oxidation in air at 900 °C, the Y-containing MAX phases exhibited lower mass gain than the undoped $Cr_2AlC$, suggesting Y has a beneficial effect on the oxidation resistance.

Lu *et al.* incorporated up to 16.7 at.% Y in the MAX phase and formed the *i*-MAX phase $(Cr_{2/3},Y_{1/3})_2AlC$ [14]. However, this *i*-MAX phase did not crystallize in the conventional $P6_3/mmc$ space group, but rather in the orthorhombic *Cmcm* and *C/2c*. ElMelegy *et al.* investigated the behavior of the *i*-MAX during oxidation in air at 1000

°C and showed the formation of a dense and protective $Cr_2O_3$ outer layer followed by a Yttrium Aluminum Garnet (YAG) ($Y_3Al_5O_{12}$) inner layer [19].

The compositional space investigated so far in the Cr-Y-Al-C system is limited to either very low contents (<0.5 at.%) or comparatively high ones (~17 at.%). So far, large amounts of Y lead to the formation of the *i*-MAX phase which crystallizes in the *Cmcm* and *C/2c* space groups. In the present work, up to 5 at.% of Y were incorporated in $Cr_2AlC$ thin films and the resulting films were analyzed with respect to their composition, structure and microstructure. To identify the mechanisms involved in the incorporation of Y in the MAX phase two approaches were used: a two-step synthesis based on co-sputtering at room temperature followed by annealing, and a direct synthesis based on co-sputtering of $Cr_2AlC$ and Y at high temperature. The phases formed were identified for each synthesis route and the composition-dependent microstructures were revealed.

## 2. Methods

### 2.1. Computational details

All first principles calculated energies were obtained using the Vienna Ab initio Simulation Package (VASP 5.4.4) [22] implementation of density functional theory (DFT), using the Perdew–Burke–Ernzerhof (PBE) generalize gradient approximation [23] description of the exchange-correlation energy. The plane wave energy cutoff was set at 400 eV, *k*-point grids with a spacing of 0.05 $Å^{-1}$ according to the Monkhorst-Pack method [24]. The electronic energy convergence threshold was set to $10^{-6}$ eV/atom for energy and $10^{-2}$ eV/Å for force.

The Special Quasi-random Structures (SQS) method [25], as implemented in the Alloy Theoretic Automated Toolkit (ATAT) package [26], was used to generate representative supercell structures that approximate a fully random alloy of Cr and Y.

The SQS supercell thus represents the best possible periodic supercell that mimics the local pair and multisite correlation functions of a random alloy under the constraint of a given supercell size *N*. Figure S1 shows the enthalpy convergence of formation enthalpy with respect to number of atoms in the supercell for selected values of *x* in $(Cr_{1-x}Y_x)_2AlC$.

The thermodynamic stability of $(Cr_{1-x}Y_x)_2AlC$ phases was investigated at 0 K with respect to decomposition into any combination of competing phases. To identify the set of most competing phases, the equilibrium simplex, a linear optimization procedure based on the simplex method was used under the constraint of a fixed MAX stoichiometry [27,28]. The stability of a phase is quantified in terms of formation enthalpy $\Delta H$ by comparing its energy to the energy of the equilibrium simplex,

$$\Delta H = E(\text{compound}) - E(\text{equilibrium simplex}). \quad (1)$$

where $\Delta H < 0$ indicates a stable phase, while $\Delta H > 0$ is considered to be not stable or at best metastable. The selection of competing phases includes all known elemental, binary, ternary, and quaternary phases within herein considered quaternary systems. Hypothetical phases were also included, based on compounds that exist in similar systems and/or with neighboring elements in the periodic table, as competing phases. A complete list of competing phases considered herein are found in Table S1.

Contributions from other temperature-dependent effects, *e.g.*, lattice vibrations and electronic entropy, to the formation enthalpy are approximated to be negligible as these, significant or not, tend to be cancelled out in the Gibbs free energy of formation term [29]. This approach has been proven to work exceptionally well for previous theoretical studies of both ternary and quaternary MAX phases [28,30–36].

However, since we are investigating a solid solution of Cr and Y, approximated through modelled disorder (SQS), contribution from configurational entropy to the Gibbs free energy of formation *G* is approximated using

$$G = H - T\Delta S, \quad (2)$$

where *T* is the temperature and $\Delta S$ the configurational entropic contribution per *M*-site, assuming an ideal solution of Cr and Y on the *M*-sites, and it is given by

$$\Delta S = -w\mathrm{k}_B[x\ln(x) + (1-x)\ln(1-x)], \quad (3)$$

where *x* is the concentration of Y.

Lattice constants were calculated within the framework of density functional theory [37,38] as implemented in the Vienna ab initio package (VASP) [22,39]. The exchange-correlation energy was calculated with the generalized gradient approximation (GGA) of the Perdew-Burke-Ernzerhof (PBE) scheme [23]. A k-point mesh of 4×7×2 was automatically constructed following the method of Monkhorst-Pack [24], and a cut-off energy of 500 eV was employed. $Cr_2AlC$ was considered in the most stable antiferromagnetic (AFM) ordering as reported by [40]. In addition to the spin polarization, a Hubbard U correction of 2 eV Cr was used [41]. 4×4×2 supercells were constructed, and the Y atoms were distributed in the Cr sublattice, equally replacing spin up/down states, following the special quasirandom structure approach as implemented in *sqsgen* [42].

## 2.2. Experimental details

$Cr_2AlC$ and Y-containing $Cr_2AlC$ films were deposited by co-sputtering from a powder metallurgical composite Cr-Al-C target (Ø 20 mm) with MAX phase stoichiometry (2:1:1) provided by Plansee Composite Materials GmbH, Germany, and an elemental Y target (Ø 20 mm, 99.5 % purity) provided by MaTeck, Germany. The Cr-Al-C target was operated in direct current magnetron sputtering (DCMS), while the Y target was operated in high power pulsed magnetron sputtering (HPPMS) to promote densification of the films [43,44]. The depositions took place in an ultra-high vacuum (UHV) chamber, with a base pressure of $5\times10^{-5}$ Pa, a working pressure of 0.55 Pa in presence of Ar (99.9999%), and a target to substrate distance of 100 mm. The geometry of the chamber is described elsewhere [45]. Depositions were carried out on $10\times10$ mm$^2$ MgO(100) substrates (Crystal GmbH, Germany) and on NaCl covered aluminum foil.

To assess whether the formation of the Y-containing MAX phase is driven by surface and/or bulk diffusion, depositions were carried out at high (650 °C) and room temperature (*i.e.* without intentional heating). In order to vary the Y content, the power density on the Y target was kept constant at 2.5 W·cm$^{-2}$, while the power density at the Cr-Al-C target was 10 and 6 W·cm$^{-2}$, for ~3 and ~5 at.% Y, respectively. During deposition of the pure $Cr_2AlC$ MAX phase, the power density applied was also 10 W·cm$^{-2}$. To ensure the homogeneity of the deposited films, the substrates were rotated during deposition. In order to obtain similar thicknesses (1-2 µm) for all films, the deposition times were 90 and 160 min, for the highest and lowest power density at the Cr-Al-C target, respectively. The HPPMS parameters applied to the Y target are provided in Table 1.

Table 1: HPPMS parameters applied to the Y target

| **Deposition Parameters** | **Y target** |
|---|---|
| Power (W) | 50 |
| on/off time (µs) | 50/9950 |
| Duty cycle (%) | 0.5 |
| Peak current density (A/cm$^2$) | ~ 2.0 |

The chemical composition of the as deposited films was measured by ion beam analysis using a 5 MeV Tandem Accelerator at Uppsala University [46]. Time-of-flight elastic recoil detection analysis (ToF-ERDA) and elastic backscattering spectrometry (EBS) were combined in order to enhance the measurement accuracy [47]. For ToF-ERDA, a primary beam of 36 MeV $^{127}I^{8+}$ was used and recoils were detected at a forward angle of 45° with incidence angle of the primary beam and exit angle of the recoils both at 22.5°. Time-energy coincidence spectra were converted into depth profiles with CONTES [48]. The ToF-ERDA depth profiles were homogeneous and O as well as F impurities of < 1 at.% were detected. For EBS, a 4.5 MeV $^{4}He^{+}$ beam was used and backscattered ions were detected at an angle of 170°. The C content in the near surface region of the films was determined through the $^{12}C(^{4}He,^{4}He)^{12}C$ elastic resonance at ~4.260 MeV [49]. SIMNRA [50] was employed for EBS data analysis. The total maximum measurement uncertainty was < 4% relative of the deduced values and an EBS spectrum recorded for as deposited $Cr_2AlC$ with 5.4 at.% Y is provided in Figure S2.

The chemical composition of the films annealed after deposition, was measured in a JEOL JSM 6480 scanning electron microscope equipped with an EDAX Genesis 2000 energy dispersive X-ray spectroscopy detector (EDX). Measurements were carried out with an acceleration voltage of 12 kV. The spectra were quantified based on a reference sample which had been measured by ion beam analysis ($Cr_{47.9}Y_{3.5}Al_{24.3}C_{24.3}$).

Phase identification was carried out in a Siemens D5000 system (Cu α radiation), in Bragg-Brentano configuration. *In situ* XRD during heating was carried out in a Bruker AXS D8 Discover X-Ray diffractometer with an integrated General Area Detector Diffraction System (GADDS), equipped with an Anton Paar DHS 1100 heating stage, using Cu α radiation. The measurements were carried out in vacuum (<6×$10^{-2}$ mbar). An incident angle of 12° was used while the measurements were carried out over a $2\theta$ range of 12-70°, with a frame width of 30° and a collection time of 5 min per frame. The samples were annealed from 500 to 700 °C with a step size of 50 °C. To ensure temperature stability, the measurements were started after 5 min of holding time after the desired temperature had been reached.

The crystallization kinetics of Cr-Al-C and Cr-Y-Al-C powders were investigated by differential scanning calorimetry (DSC, Netzsch Jupiter® STA 449 C). The films deposited on the NaCl-covered aluminum foil were separated by dissolving the NaCl in deionized water. The resulting flakes were then filtered, cleaned in distilled water and acetone, and dried. The DSC measurements were performed in 30 sccm flow of Ar (99.9999%) in continuous heating mode. The powder samples were first pre-heated at 150 °C for 30 min for outgassing and removal of organic impurities. An integrated oxygen trap system (OTS®) enabled a low oxygen concentration (<1 ppm) in the measurement atmosphere. The powders were then heated to 1200 °C with heating and cooling rates of 10 and 40 °C/min, respectively. Approximately 20 mg were used for each measurement. The remaining Cr-Al-C and Cr-Y-Al-C powders were annealed to 750 and 760 °C, respectively, with a heating rate of 10 °C/min. The powder samples were then sealed in fused silica capillaries with a wall thickness of 20 μm and a diameter of 1 mm under Ar (99.9999% purity) atmosphere.

High-energy XRD measurements on selected powdered coatings were performed in a transmission geometry at the beamline P02.1 [51] of Deutsches Elektronen Synchrotron (DESY) in Hamburg, Germany utilizing X-rays with a wavelength of 0.20701 Å.

Additional annealing experiments of films were carried out using a horizontal Nabertherm tube furnace. To ensure a low-pressure atmosphere, a turbomolecular pump was attached to the annealing setup. The average pressure during the annealing was $1\times10^{-6}$ mbar. The pressure was monitored during the annealing process and remained constant. The annealing program consisted of heating the samples to 760 °C and subsequent cooling to room temperature. A heating/cooling rate of 10 °C /min was used for both processes.

The microstructure was investigated using scanning transmission electron microscopy (STEM) on thin lamellae which were prepared by focused ion beam (FIB), with $Ga^+$ ions accelerated at 30 kV, in a FEI Helios NanoLab dual-beam microscope. A STEM III detector was used in high-angle annular dark field (HAADF) mode for imaging with a voltage and current of 30 kV and 50 pA, respectively. EDX linescans were collected using an EDAX Octane Elect EDX detector, using an accelerating voltage of 12 kV and 50 nm step size (standard-less quantification).

## 3. Results and discussion

Solid solutions of Cr and Y have been modelled using the special quasi-random structures (SQS) method. Figure 1(a),(b) shows the calculated stability for $(Cr_{1-x}Y_x)_2AlC$ MAX phases with respect to competing phases. See Ref. [35,36,52,53] for examples of thermal stability calculated for other quaternary MAX phases with solid solution disorder on the *M*-site. The stability is given by the formation enthalpy $\Delta H$, calculated at 0 K. Already beyond 2.5 at.% Y (x = 0.05) on the M-site is found to not be stable, as shown in Figure 1(a). Contribution from configurational entropy to the Gibbs free energy due to disorder have been estimated at 1000 and 2000 K and we find only a minor impact from an increased temperature since $\Delta H$ increases with increasing Y content. Note that in Figure 1(a), $Cr_2AlC$ has been excluded as a competing phase and the identified set of most competing phases are a combination of $Cr_2Al$, $Cr_3C_2$, $YAl_3C_3$ and $Al_4C_3$. If $Cr_2AlC$ is considered as a competing phase (Figure 1(b)) the situation becomes different as $(Cr_{1-x}Y_x)_2AlC$ is found not to be stable for $x > 0$ and should instead decompose into $Cr_2AlC$, $YAl_2$, $Y_2Cr_2C_3$ and $Cr_7C_3$.

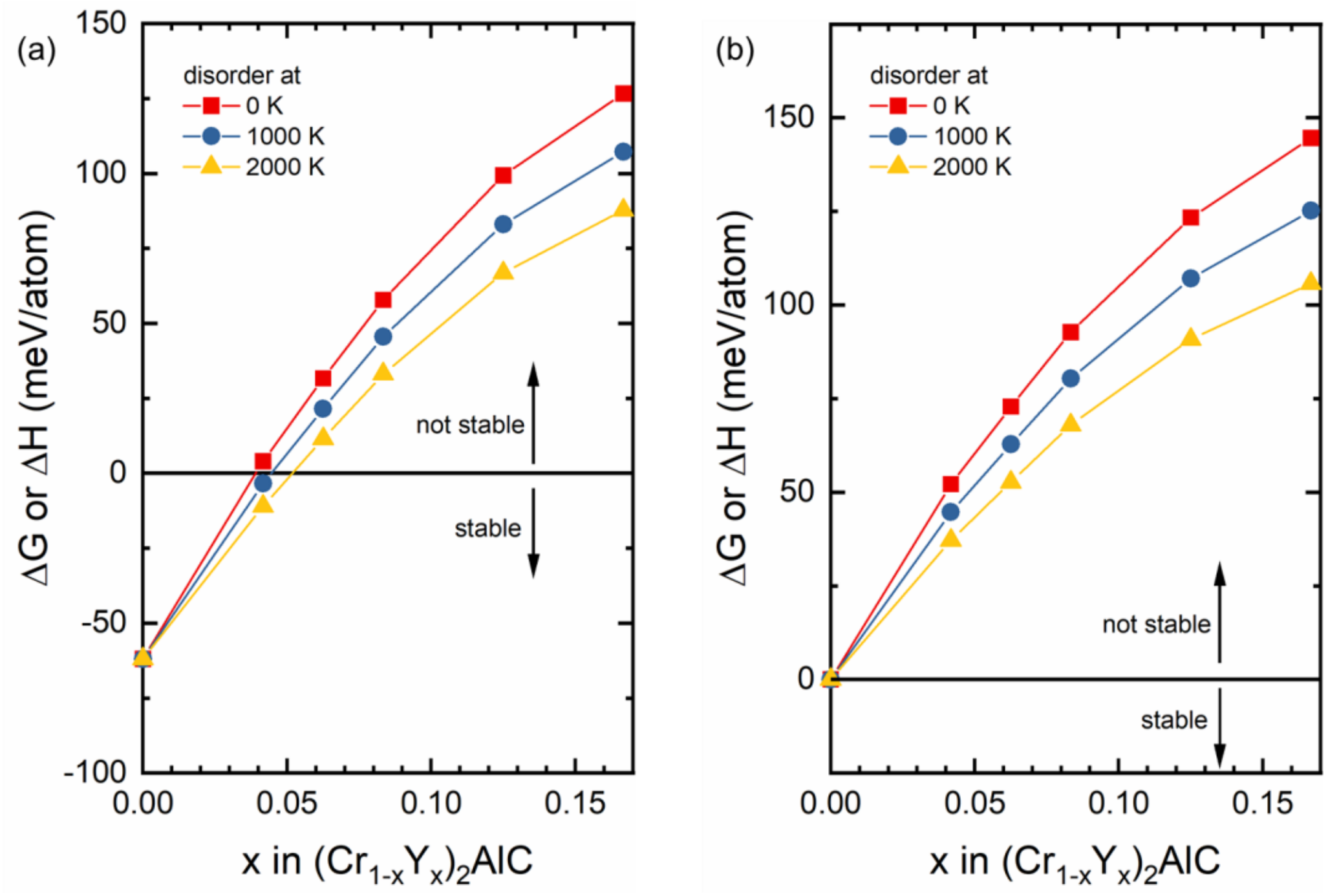

Figure 1: Calculated stability for disordered solid solution for $(Cr_{1-x}Y_x)_2AlC$ MAX phases when (a) $Cr_2AlC$ is not considered as competing phase and (b) with $Cr_2AlC$ considered as a competing phase.

The chemical compositions of the films deposited without intentional heating and at high temperature (650 °C) are reported in Table 2. Oxygen contents were ≤ 0.8 at.% for all films, however, the incorporated oxygen concentrations were slightly higher in presence of Y, which may be due to oxygen incorporation in the Y target. In addition, fluorine impurities ≤ 0.6 at.% were detected in some of the Y-containing as deposited films with no intentional heating. Both impurities will not be considered in the following discussions. The compositions of the films were unaffected by heating as is evidenced by the small deviations between 650 °C and room temperature depositions. For the sake of clarity, the samples will be referred to as $Cr_2AlC$, $Cr_2AlC$ + 3 at.% Y and $Cr_2AlC$ + 5 at.% Y.

Table 2: Chemical composition of films deposited with no intentional heating and at 650 °C, determined by EBS and EBS-corrected EDX.

| Samples | Cr (at.%) | Y (at.%) | Al (at.%) | C (at.%) | O (at.%) | F (at.%) |
|---|---|---|---|---|---|---|
| **As deposited with no intentional heating** | | | | | | |
| **Cr-Al-C** | 48.5 ± 1.8 | - | 25.9 ± 0.9 | 25.3 ± 0.9 | 0.3 ± 0.1 | - |
| **Cr-Y-Al-C (10 W·cm$^{-2}$)** | 47.5 ± 1.7 | 3.5 ± 0.1 | 24.1 ± 0.9 | 24.1 ± 0.9 | 0.4 ± 0.1 | 0.4 ± 0.1 |
| **Cr-Y-Al-C (6 W·cm$^{-2}$)** | 46.4 ± 1.7 | 5.4 ± 0.2 | 23.8 ± 0.9 | 23.3 ± 0.9 | 0.5 ± 0.1 | 0.6 ± 0.1 |
| **As deposited at 650 °C** | | | | | | |
| **Cr-Al-C** | 48.5 ± 1.8 | - | 25.1 ± 0.9 | 26.0 ± 0.9 | 0.4 ± 0.1 | - |
| **Cr-Y-Al-C (10 W·cm$^{-2}$)*** | 48.8 ± 1.5 | 3.2 ± 1.5 | 26.2 ± 1.5 | 21.2 ± 1.6 | 0.5 ± 0.1 | - |
| **Cr-Y-Al-C (6 W·cm$^{-2}$)** | 46.7 ± 1.7 | 5.7 ± 0.2 | 23.8 ± 0.9 | 23.0 ± 0.8 | 0.8 ± 0.1 | - |

*EBS-standard corrected EDX

The films deposited without intentional heating were sputtered onto MgO substrates and NaCl covered Aluminium foil. The powders collected after dissolution of the NaCl layer were measured by DSC in order to identify their crystallization conditions. The DSC data obtained for $Cr_2AlC$ and $Cr_2AlC$ +5 at.% Y deposited without intentional heating are shown in Figure 2. The DSC signal of the $Cr_2AlC$ powder is in very good agreement with data reported by Abdulkadhim *et al.* [54]. The first exothermic peak at ~558 °C corresponds to the formation of the disordered hexagonal $(Cr,Al)_2C_x$ solid solution, while the second peak at ~593 °C corresponds to the formation of the $Cr_2AlC$ MAX phase. Interestingly, the Y-containing powder exhibits a different behavior. In fact, three exothermic peaks occurring at ~757, ~770, and ~ 824 °C are identified. Furthermore, it must be noted that the first peak occurs at a temperature which is 197.5 °C higher than that of the first peak observed for the pure MAX phase, indicating that the activation energy to crystallize the Cr-Y-Al-C powder is substantially higher than that of $(Cr,Al)_2C_x$ and $Cr_2AlC$.

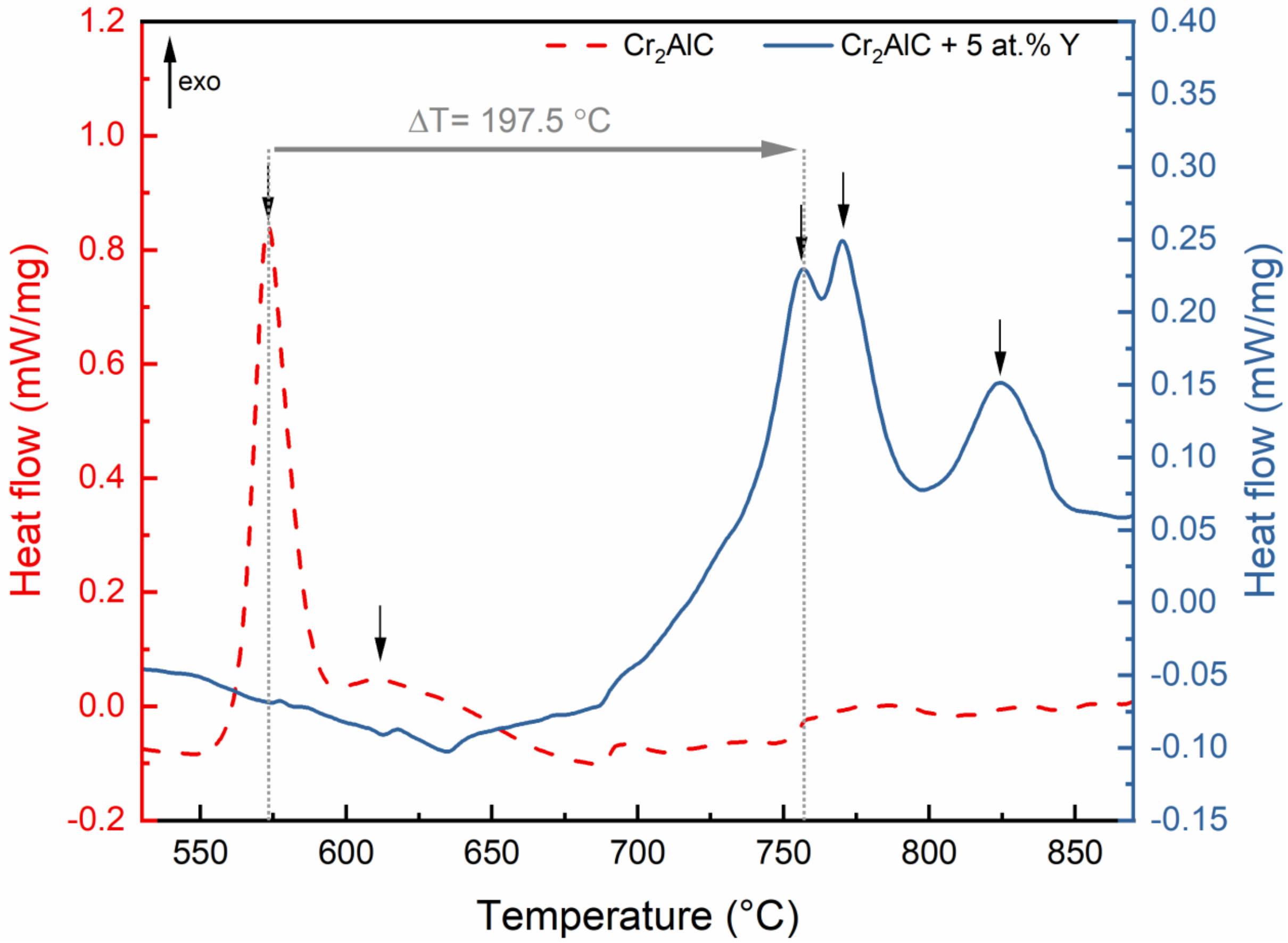

Figure 2: DSC measurements on $Cr_2AlC$ (in red dashed, left scale) and $Cr_2AlC$+ 5 at.% Y (in blue, right scale) powders with a heating rate of 10 °C/min. The heat flow data are obtained from the difference between the first and second heating cycle. The black arrows indicate the maximum heat flow for the observed exothermic reactions.

The phase formation was further investigated by annealing $Cr_2AlC$ + 5 at.% Y films in the DSC and stopping the heating sequences within the three peak temperatures of 760, 810 and 910 °C, see Figure 3(a). The 760 °C exothermic peak corresponds to MAX phase formation. It can be noted that the diffraction peaks are shifted to lower $2\theta$ values. Indeed, the original MAX phase peak positions are provided in Figure 3(b) grey solid lines (ICDD: 00-029-0017), whereas the observed peak positions of the $(Cr,Y)_2AlC$ solid solution, are provided in black dashed lines on the same figure. The XRD pattern of the sample annealed to 810 °C shows a decreased intensity of the MAX phase solid solution peaks, as well as the formation of $Cr_7C_3$ (00-036-1482), while at 910 °C the MAX phase peaks are shifted back to the positions corresponding to pure $Cr_2AlC$ and are accompanied by the appearance of peaks belonging to $Cr_3C_2$ (00-035-0804) and $Y_5Al_3$ (00-039-0775). These observations suggest that the Y-containing MAX phase is metastable and starts decomposing at ~810 °C. During the decomposition, Y is expulsed from the MAX phase causing it to form an aluminide, while the Cr and C surpluses form carbides. These observations are consistent with the quantum mechanical predictions as the competing phases included $Cr_2AlC$, $YAl_2$, $Y_2Cr_2C_3$ and $Cr_7C_3$. While $Y_5Al_3$ was indexed instead of $YAl_2$ in Figure 3(b), $Y_5Al_3$ is known to be metastable and decompose into stable phases, including $YAl_2$ [55].

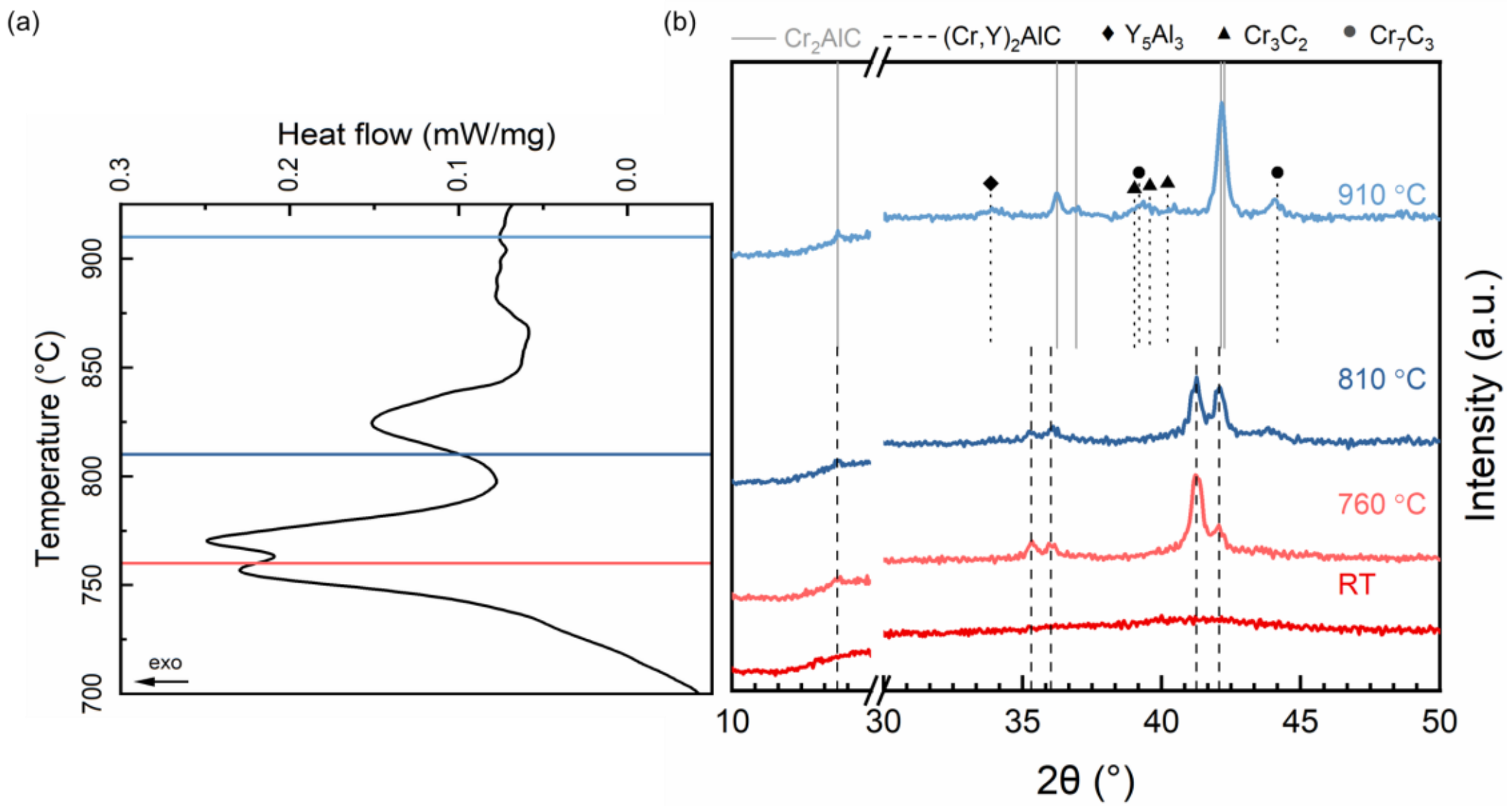

Figure 3: (a) DSC measurement of $Cr_2AlC$ + 5 at.% Y powders. The colored lines correspond to the temperatures at which the samples were annealed to and (b) corresponding XRD patterns.

To confirm the peak shift observed after annealing to 760 °C, $Cr_2AlC$+ 5 at.% Y powders were annealed to 760 °C and were then ground and mounted in a capillary for high energy X-ray powder diffraction (HEXRD) at DESY. Pure $Cr_2AlC$ powder was also annealed at 750 °C, ground and used as reference. The diffraction patterns obtained are given in Figure 4. Peak identification was carried out using the ICDD: 00-029-0017, in red dashed lines. Both powders were quasi-phase pure as a peak at ~44°, which could correspond to $Cr_7C_3$ can be seen. Considering that 5 at.% of Y corresponds to x= 0.10, the shaded areas are in good agreement with the experimental data presented in Figure 4. The $Cr_2AlC$ MAX phase is identified in both cases, however, the $Cr_2AlC$ + 5 at.% Y peaks exhibit a distinct shift to lower angles. This shift, however, is less obvious for *(00l)* reflections as shown in Figure 4(b), (c) and (d), indicating that the expansion of the lattice may be more significant in the basal plane of the MAX phase. The expansion is further evidenced by the peak splitting observed in Figure 4(b). Indeed, the *(103)* and *(006)* reflections are typically superimposed in pure MAX phases. Therefore, the lattice expansion can potentially inform on the position of Y in the MAX phase structure.

To further describe the shift to lower angles, the simulations discussed previously were used to identify the lattice parameters of the $(Cr_{1-x}Y_x)_2AlC$ phase for different x contents and are shown in Figure 5. The blue lines in Figure 4(b), (c), (d) and (e) correspond to the peak positions describing the $(Cr_{1-x}Y_x)_2AlC$ phase for x = 0.10, simulated from the theoretical lattice parameters in Figure 5. A linear regression analysis was used on both a and c datasets in order to deduce the lattice parameters at x = 0.10. The simulated peak positions were determined using VESTA.

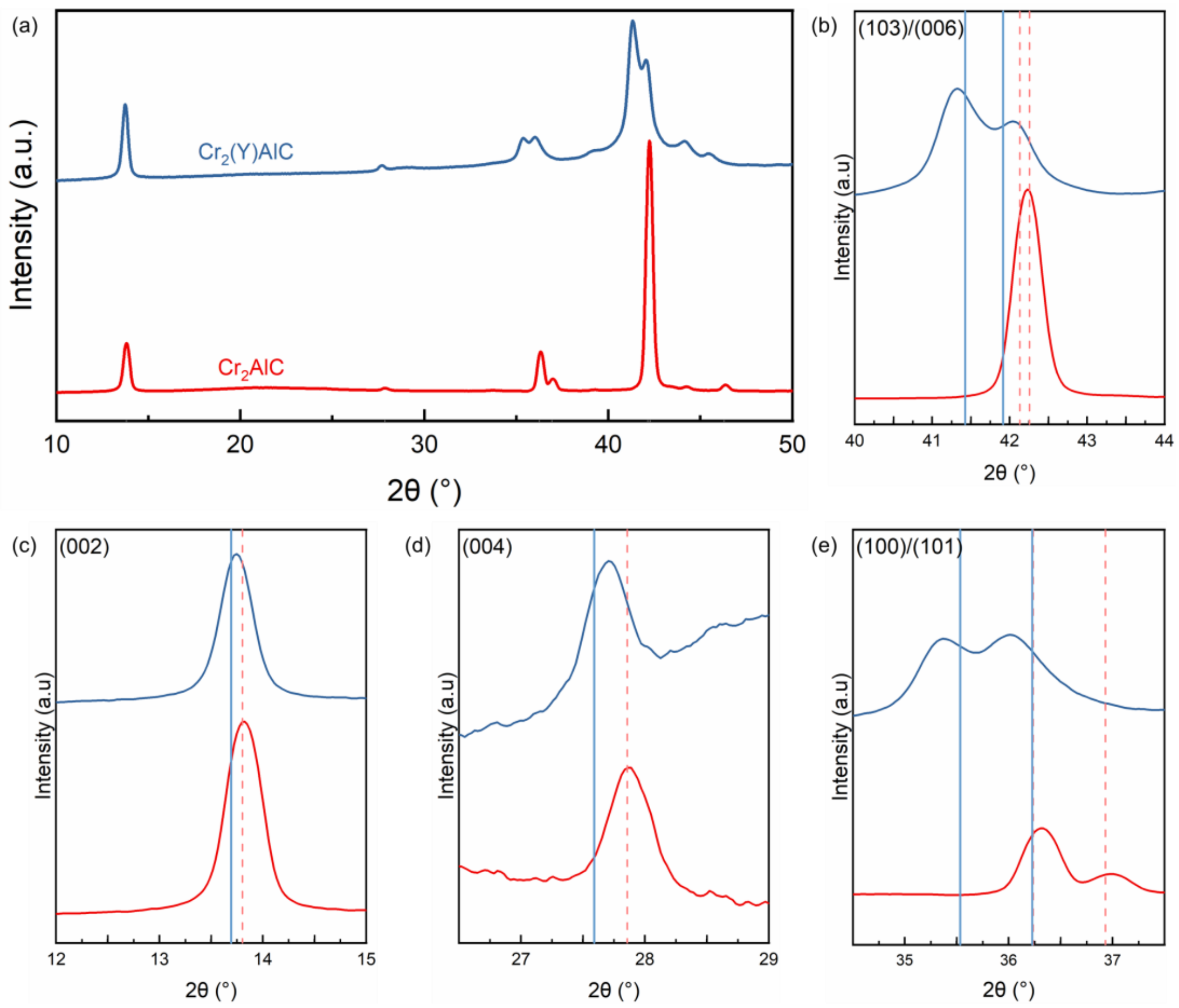

Figure 4: Synchrotron powder diffraction data from $Cr_2AlC$ and $Cr_2AlC$ + 5 at.% Y powders annealed to 750 and 760 °C, respectively, in the DSC. Black circle could correspond to $Cr_7C_3$. Detailed views of the peak shifts induced by the presence of Y are provided in (b), (c), (d), (e). The red dashed lines correspond to the peak positions based on ICDD: 00-029-0017. The blue lines represent simulated peak positions, extrapolated from the lattice parameter calculations, assuming a linear regression.

The calculated lattice parameters *a* and *c* for $(Cr_{1-x}Y_x)_2AlC$ MAX shown in Figure 5, were calculated without magnetism and were found to increase with increasing Y content. This is expected since Y is a larger atom than Cr. The lattice parameters extracted from the synchrotron data are also plotted in Figure 5 in red (*a* parameters) and blue (*c* parameters). As can be seen, the theoretical values are in very good agreement with the experimentally deduced parameters, as the deviation is <1% for both a and c parameters, suggesting that the Y incorporated in the MAX phase replaces Cr atoms on the M-sites.

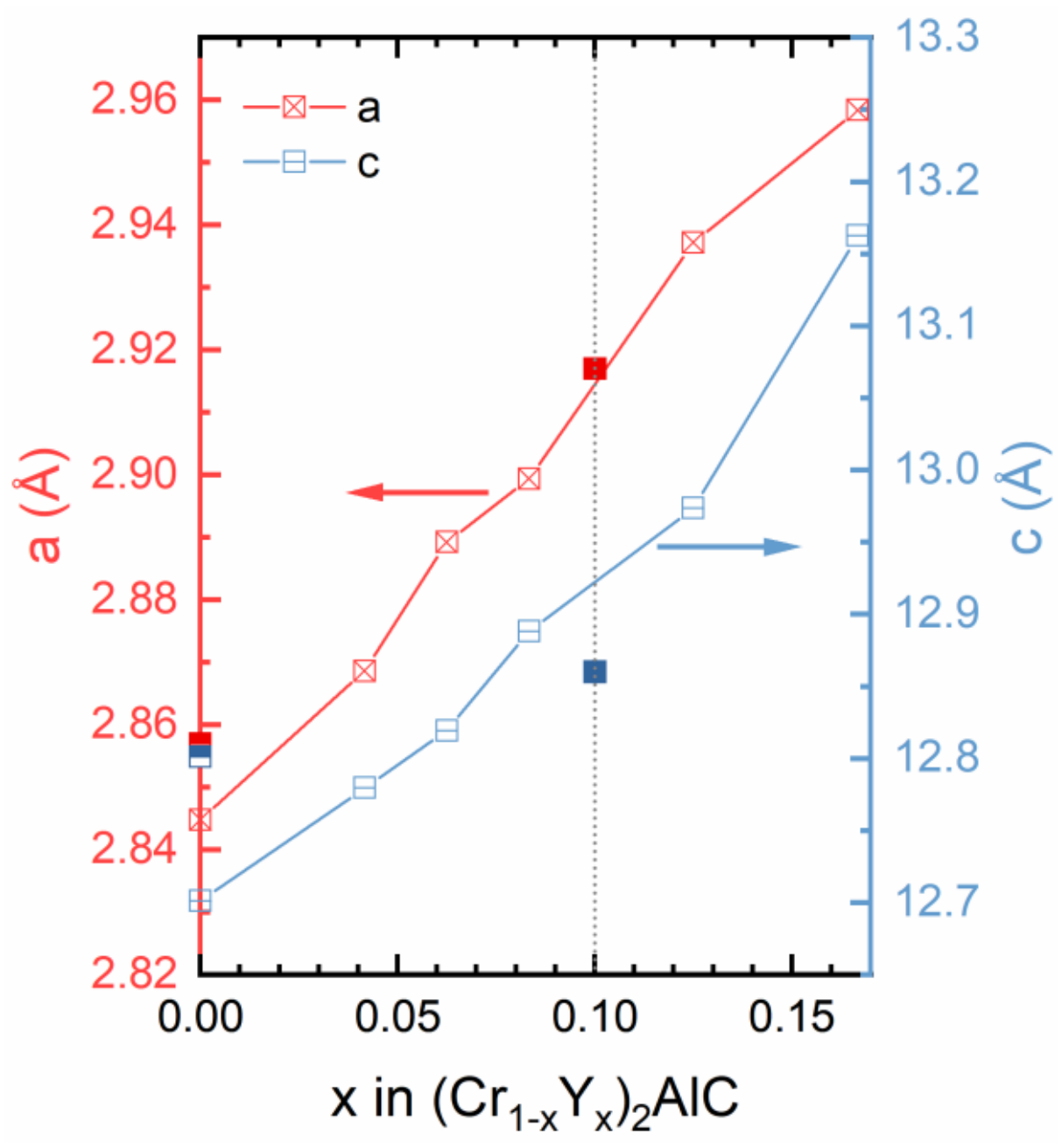

Figure 5: Calculated and experimental lattice parameters. The *a* and *c* parameters of $Cr_2AlC$ are given by the half-full squares and those of $Cr_2AlC$+5 at.% Y are given by the full squares. *a* parameters in red, *c* parameters in blue. Dashed line corresponds to 5 at.% of Y, i.e. x = 0.10.

The expansion of the lattice is observed as the extracted *a* and *c* lattice parameters of the $(Cr,Y)_2AlC$ MAX phase increase when compared to the $Cr_2AlC$ MAX phase, as the structure attempts to accommodate the large Y atoms. Therefore, based on these observations and the lack of secondary phases, one can assume that $Cr_2AlC$ can accommodate, at least, up to 5 at.% of Y in its original *$P6_3/mmc$* structure. These observations are consistent with the conventional XRD data provided in Figure 3.

Experimental and calculated lattice parameters are gathered in Table 3. Calculations were carried out without magnetism and by taking the stable antiferromagnetic (AFM) ordering of $Cr_2AlC$ into account [40]. In the case of AFM ordering, the lattice parameters are overestimated when considering M-site replacement, particularly for the Y-containing MAX phase. When calculating the lattice parameters of a MAX phase where the Y replaces Al atoms, the *a* parameter does not vary significantly from the M-site calculations, although both values are larger than the measured one. However, the *c* parameter is overestimated by more than 4% compared to the experimental data, when Al is replaced by Y. Therefore, these calculations also suggest that Y is not positioned on the Al-site but rather replaces Cr atoms on the M-site. The deviation, however, between experimental and theoretical values obtained with AFM ordering is between 1 and 3%, and therefore, much larger than the 0.1-0.5% deviation between the experimental and theoretical values without spin polarization, indicating that the consideration of magnetism, in this case, does not enhance the prediction.

Table 3: Lattice parameters extracted from synchrotron powder diffraction and calculations using the antiferromagnetic ordering of $Cr_2AlC$

| Sample | *a* parameter (Å) | *c* parameter (Å) | *Deviations a/c* |
|---|---|---|---|
| **Measured** | | | |
| **$Cr_2AlC$** | 2.857(1) | 12.801(2) | |
| **$Cr_2AlC$ + 5 at.% Y** | 2.917(3) | 12.860(8) | |
| **Calculated (no magnetism)** | | | |
| **$(Cr_{1-x}Y_x)_2AlC$ x=0.10*** | 2.915 | 12.922 | 0.1/0.5 |
| **Calculated (AFM ordering) – M-site** | | | |
| **$Cr_2AlC$** | 2.879 | 12.869 | 1.3/0.1 |
| **$Cr_2AlC$ + 5 at.% Y** | 3.012 | 13.289 | 3.1/3.2 |
| **Calculated (AFM ordering) – A-site** | | | |
| **$Cr_2AlC$ + 5 at.% Y** | 2.994 | 13.440 | 2.5/4.3 |

*In situ* XRD during vacuum annealing, was carried out on $Cr_2AlC$ and $Cr_2AlC$ + 3 at.% Y films from 500 to 700 °C, and is reported in Figure 6. The broad diffraction signal at ~62° corresponds to the signal of the graphite dome used during the measurement. A shift in crystallization temperature of ~100 °C can be noted for this Y content. The shift is almost half of that detected by the DSC for the $Cr_2AlC$ + 5 at.% Y powder, suggesting that the crystallization temperature is Y-content dependent. Contrary to the pure $Cr_2AlC$, the Y-containing MAX phase crystallizes directly into the MAX phase and not, beforehand, into a disordered solid solution, as also noted for the 5 at.% Y sample. Furthermore, although not as significant as for the 5 at.% Y powder, the peak split of the *(103)* and *(006)* reflections at ~41.5° can be observed in Figure 6(b). The lattice parameters were not extracted here as the contribution of the thermal expansion would need to be corrected for reliable results. Up to 700 °C, no decomposition of secondary phases could be detected for either of the MAX phases.

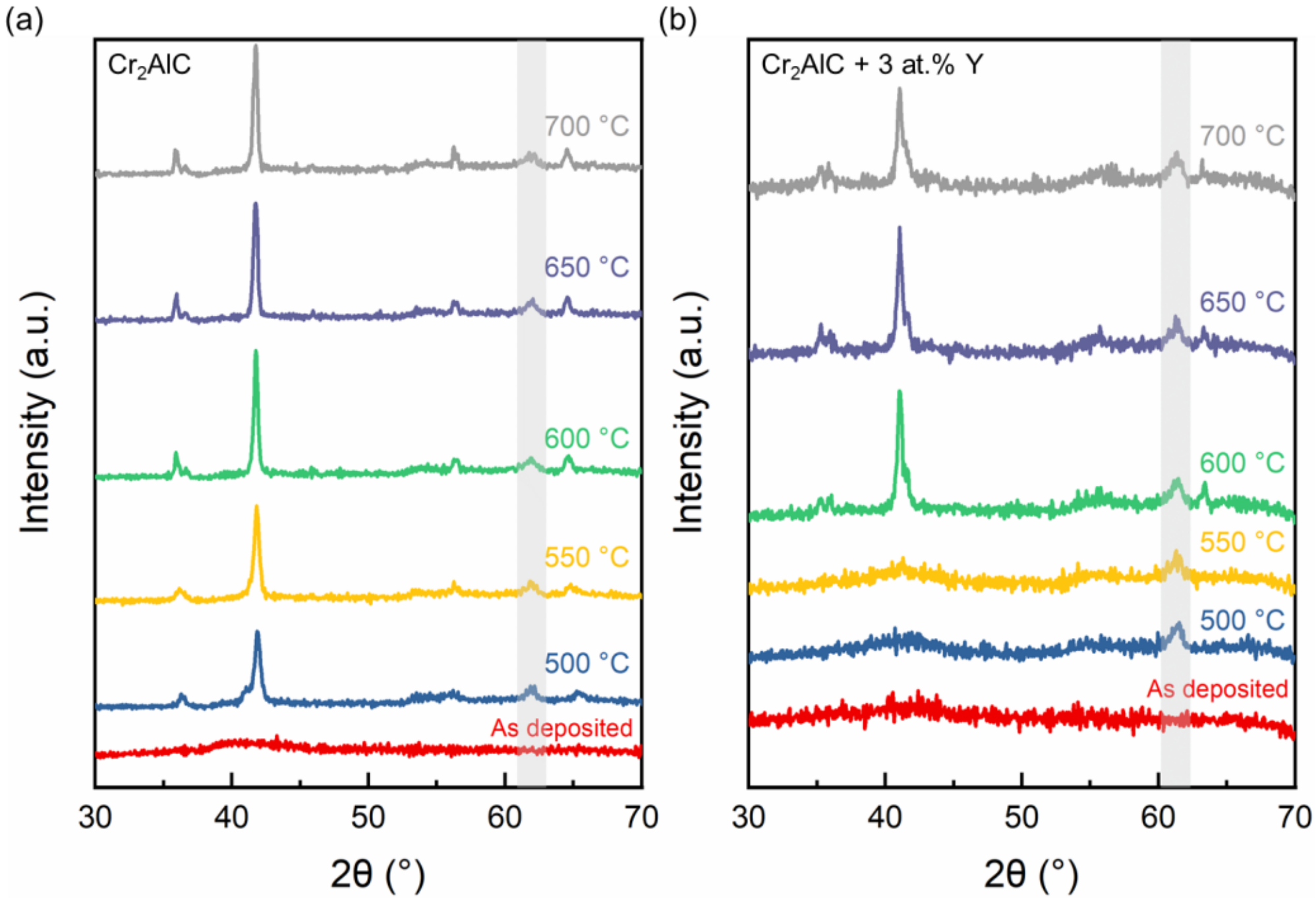

Figure 6: Patterns obtained during *in situ* XRD of (a) $Cr_2AlC$ and (b) $Cr_2AlC$ +3 at.% Y films between 500 and 700°C. Greyed band (~61°) corresponds to the signal of the graphite dome used during the *in situ* XRD measurements. All other reflections can be attributed to the $Cr_2AlC$ / $(Cr,Y)_2AlC$ MAX phase. The as deposited patterns were collected without the graphite dome, at room temperature.

The effect of Y on the microstructure of the films was evidenced by cross-section STEM imaging, presented in Figure 7. The annealed $Cr_2AlC$ film exhibits an equiaxed microstructure composed of grains of tens to hundreds of nanometers in size and a homogeneous composition, based on the EDX linescan. In the case of the Y-containing MAX phase, a different microstructure can be observed in the STEM cross-section in Figure 7(b). The grains appear much smaller, suggesting that the Y is impeding grain growth. Furthermore, the Y is well distributed throughout the thickness of the film as evidenced by EDX and the little-to-none contrast in HAADF. Hence, the Y-containing MAX phase shows potential for high temperature applications as a finer microstructure led to the formation of a passivating $Al_2O_3$ scale in the case of bulk $Ti_3AlC_2$ and would be preferred to coarser microstructures [56]. Furthermore, Chen *et al.* showed that equiaxed $Cr_2AlC$ coatings are more oxidation resistant than columnar morphologies [57].

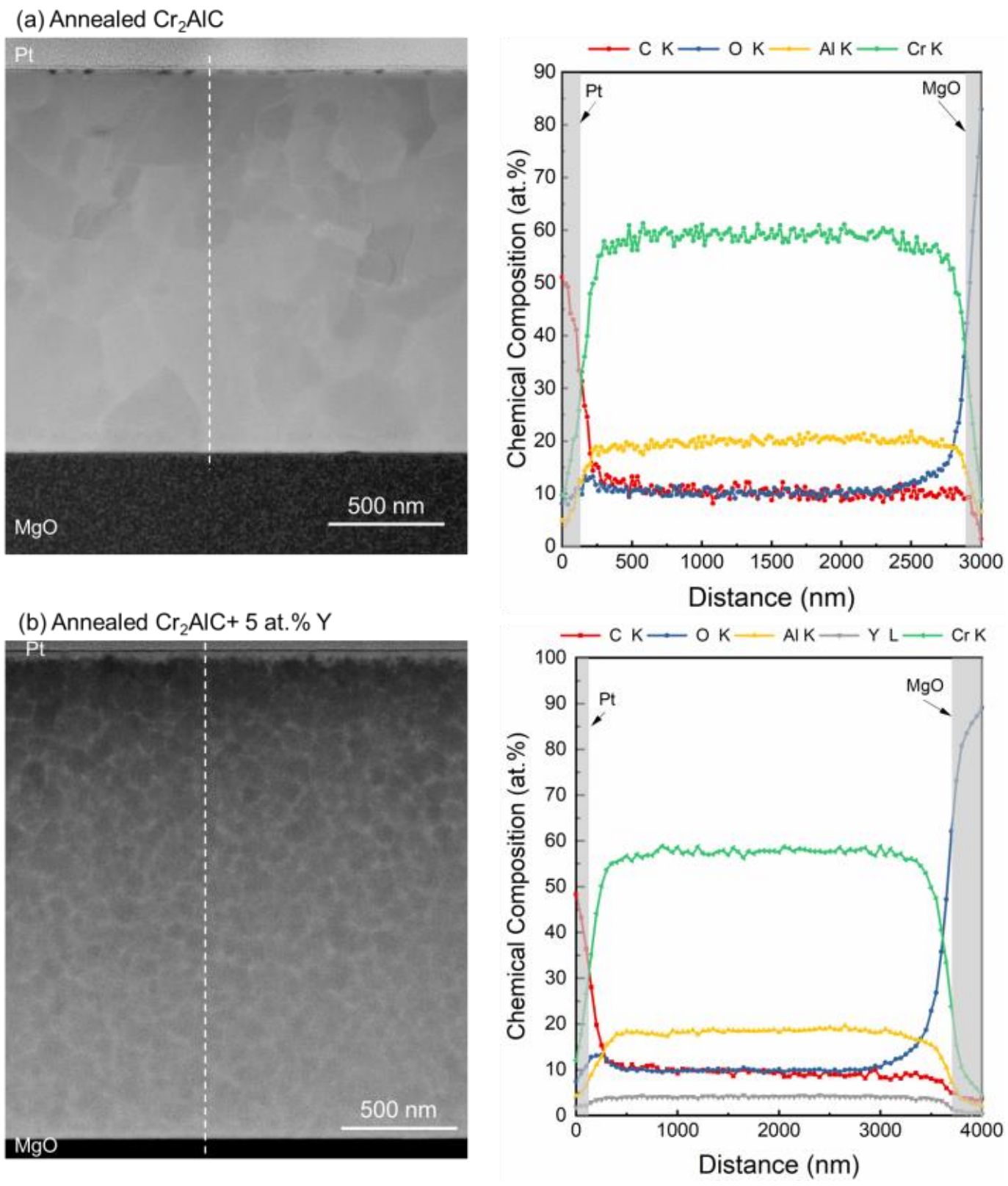

Figure 7: STEM images in HAADF mode and corresponding linescans of (a) $Cr_2AlC$ annealed at 690 °C and (b) $Cr_2AlC$+ 5 at.% Y annealed at 760 °C, in the UHV furnace with heating rates of 10 °C/min.

The XRD patterns of films deposited at 650 °C are provided in Figure 8. While the $Cr_2AlC$ films exhibits the characteristic MAX phase peaks (ICDD: 00-029-0017), Y incorporation led to the disappearance of the peak located at ~13°, corresponding to the *(002)* reflection of the MAX phase, and to the appearance of secondary phases including Cr-carbides ($Cr_3C_2$: 00-035-0804; $Cr_7C_3$: 00-036-1482; $Cr_{23}C_6$: 00-035-0783) and $Y_5Al_3$ (00-039-0775). While the MAX phase peaks at ~36°, ~37°, and ~41.5° are still observed, even for 5 at. % of Y, the secondary phases indicate the formation of multiphase films. Hence, co-sputtering of Y with Cr-Al-C under these conditions, does not lead to a single-phase Y-containing MAX phase. Since MAX phases are stoichiometry-sensitive, large amounts of Cr on the M-sites can interfere with the positioning of the larger Y atoms, leading to a surplus of Y in the form of an aluminide. Additionally, the increased surface diffusion during deposition at high temperatures impacts in particular the smaller atoms, allowing for the most stable form of the MAX phase, *i.e.* $Cr_2AlC$, to form because of increased Cr atom mobility. To counteract that effect, higher deposition temperatures and/or ion bombardment may be required in order to provide the necessary mobility to overcome the activation energy barrier associated with the formation of the solid solution. However, at higher temperatures and intense ion bombardment Al loss is observed [58,59].

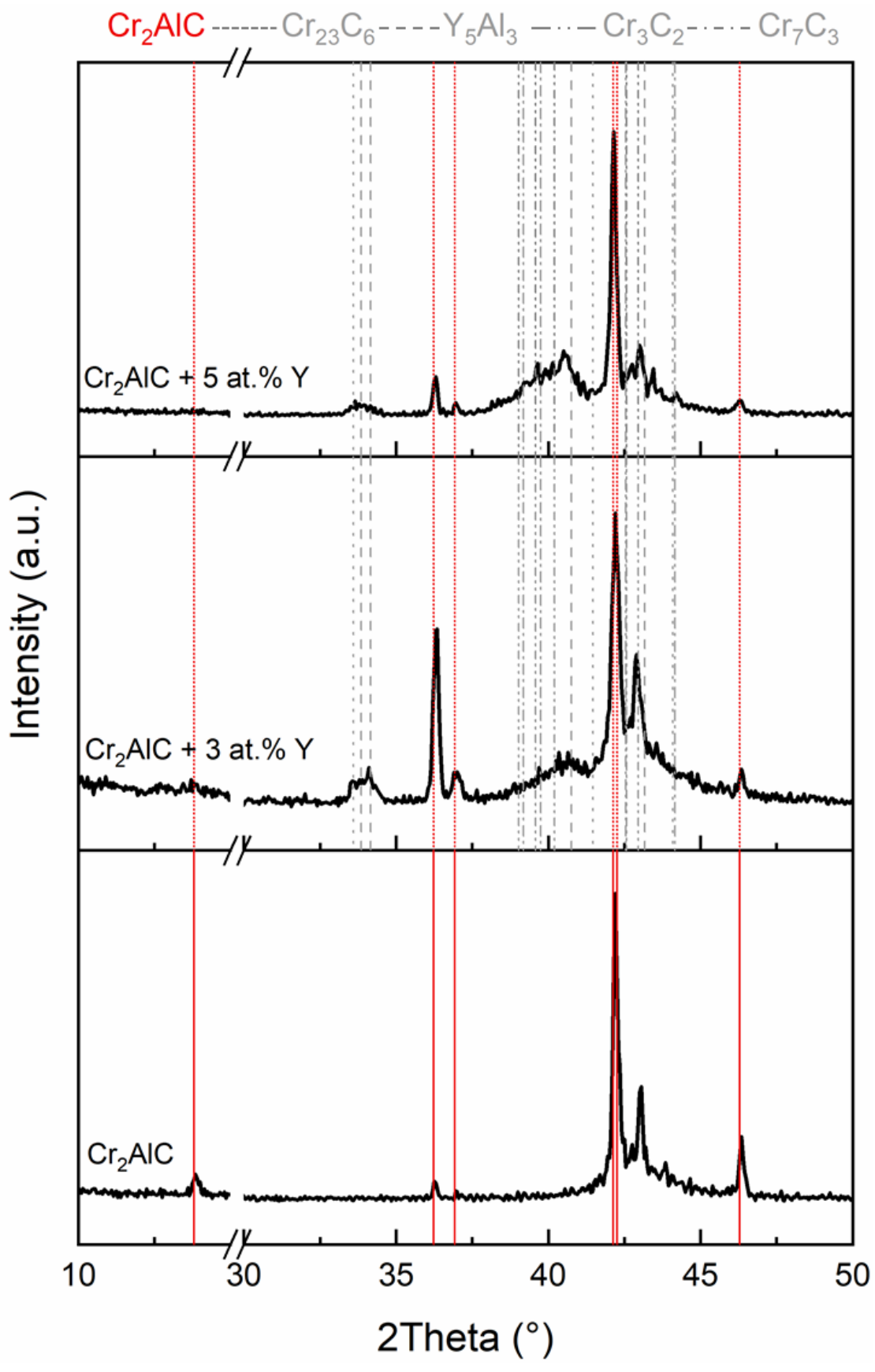

Figure 8: XRD patterns of Cr2AlC without and with Y (3 and 5 at.%), deposited at 650 °C on MgO substrates.

Surface micrographs and cross-section STEM images of the films deposited at 650 °C, along with EDX linescans are shown in Figure 9. One can notice from the surface micrographs that the $Cr_2AlC$ film is denser than the $Cr_2AlC$ +5 at.% Y films. Indeed, small inter-grain porosity can be observed on the surface. This statement is further supported by the cross-section STEM images where inter-columnar pores are visible in Figure 3(b). The distribution of Y throughout the thickness of the film appears homogeneous, although chemical contrast on the STEM image may indicate a slightly different elemental distribution.

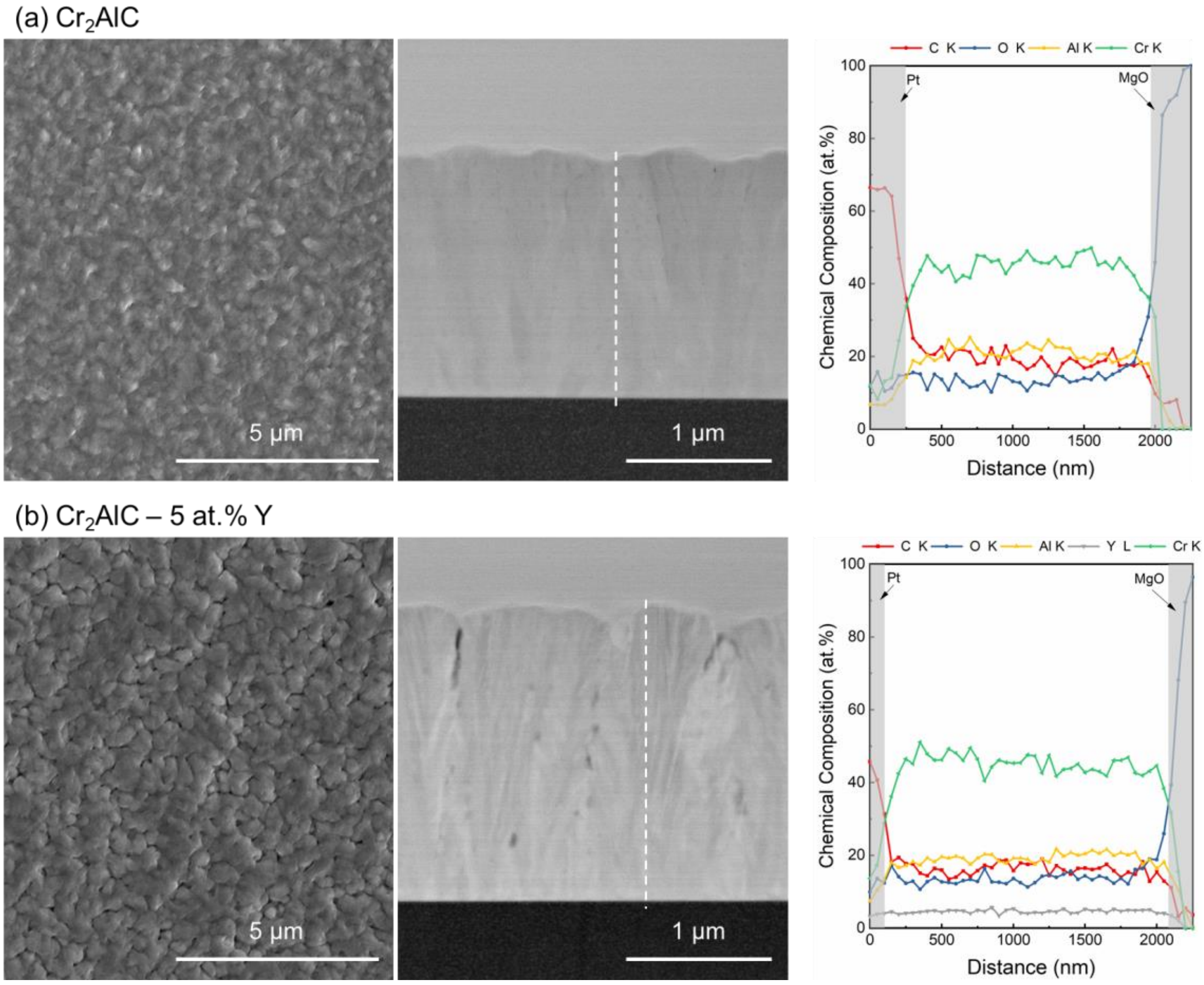

Figure 9: Surface SEM micrographs, cross-section STEM images in HAADF mode and EDX linescans of (a) $Cr_2AlC$ and (b) $Cr_2AlC$+ 5 at.% Y.

## 4. Conclusions

First-principles calculations predict the metastability of $(Cr,Y)_2AlC$ phases which was also confirmed experimentally. Y contents up to 5 at.% were explored to assess whether the formation of Y-containing $Cr_2AlC$ was possible, while maintaining the original $P6_3/mmc$ structure of the MAX phase. A two-step approach and a direct high temperature synthesis approach were carried out in order to assess the formation of the MAX phase based on surface and bulk diffusion, respectively.

The two-step synthesis involved co-sputtering of the Y and MAX phase, followed by annealing. DSC measurements allowed identifying three reactions starting at 710 °C, instead of the two reactions at 558 and 593 °C describing the crystallization of pure $Cr_2AlC$. Combined with *in situ* XRD during annealing, the dependence of the crystallization temperature on the Y content was demonstrated.

Up to 5 at.% of Y was incorporated in the $P6_3/mmc$ MAX phase and high energy X-ray powder diffraction allowed retrieving the lattice parameters of the $(Cr,Y)_2AlC$ phase. The effect of Y incorporation on the lattice was evidenced by the increase in lattice parameters which were in very good agreement with calculated lattice parameters. Furthermore, DFT predictions suggested that Y would rather replace the metal on the M-site, *i.e.* Cr, rather than the A-site *i.e.* Al, and that the system was not spin polarized. Furthermore, Y caused grain refinement, as evidenced by a significant decrease of the grain size when compared to the pure $Cr_2AlC$ and was well distributed throughout the film as no clear indication of Y segregation was found by STEM/EDX.

Finally, the high temperature synthesis led to the formation of multiphase films exhibiting under-dense column boundaries for Y contents up to 5 at.%. The formation of such films is likely due to off-stoichiometry effects and to the deposition parameters which did not allow overcoming the activation energy barrier necessary for the formation of the $(Cr,Y)_2AlC$ MAX phase.

**Acknowledgements**

This project has received funding from the European Union’s Horizon 2020 research and innovation programme under the Marie Skłodowska-Curie grant agreement No. 892501 (REALMAX). M.D and J. R. acknowledges funding from the Swedish Foundation for Strategic Research (SSF), EM16-0004. The calculations were carried out using supercomputer resources provided by the Swedish National Infrastructure for Computing (SNIC) at the National Supercomputer Centre (NSC) partially funded by the Swedish Research Council through grant agreement no. 2018-05973. The authors also gratefully acknowledge the computing time granted by the JARA Vergabegremium and provided on the JARA Partition part of the supercomputer CLAIX at RWTH Aachen University (project JARA0221). Transnational access to the ion beam analysis facility at Uppsala University has been supported by the RADIATE project under the Grant Agreement 824096 from the EU Research and Innovation program HORIZON 2020. Accelerator operation at Uppsala University has been supported by the Swedish research council VR-RFI (#2019-00191). The authors acknowledge DESY (Hamburg, Germany), a member of the Helmholtz Association HGF, for the provision of experimental facilities. The authors acknowledge DESY (Hamburg, Germany), a member of the Helmholtz Association HGF, for the provision of experimental facilities. Parts of this research were carried out at PETRA III beamline P02.1

**Declaration of Competing Interest**

The authors declare that they have no known competing financial interests or personal relationships that could have appeared to influence the work reported in this paper.

**Author Contributions**

C.A. and T.B. conceived the research. T.B. synthesized the samples, performed annealing experiments, EDX and XRD measurements. C.A. prepared the thin lamellae and performed the STEM imaging followed by EDX. D.M.H. performed DSC measurements. M.H. and D.P. performed ion beam analysis. A.S.J.M. performed the powder diffraction measurements at DESY. Calculations were carried out and interpreted by M.D., J.R., and L.L.. C.A. and J.M.S. have supervised the project. The manuscript was primarily written by C.A. with input from all authors.

## Supplemental Information

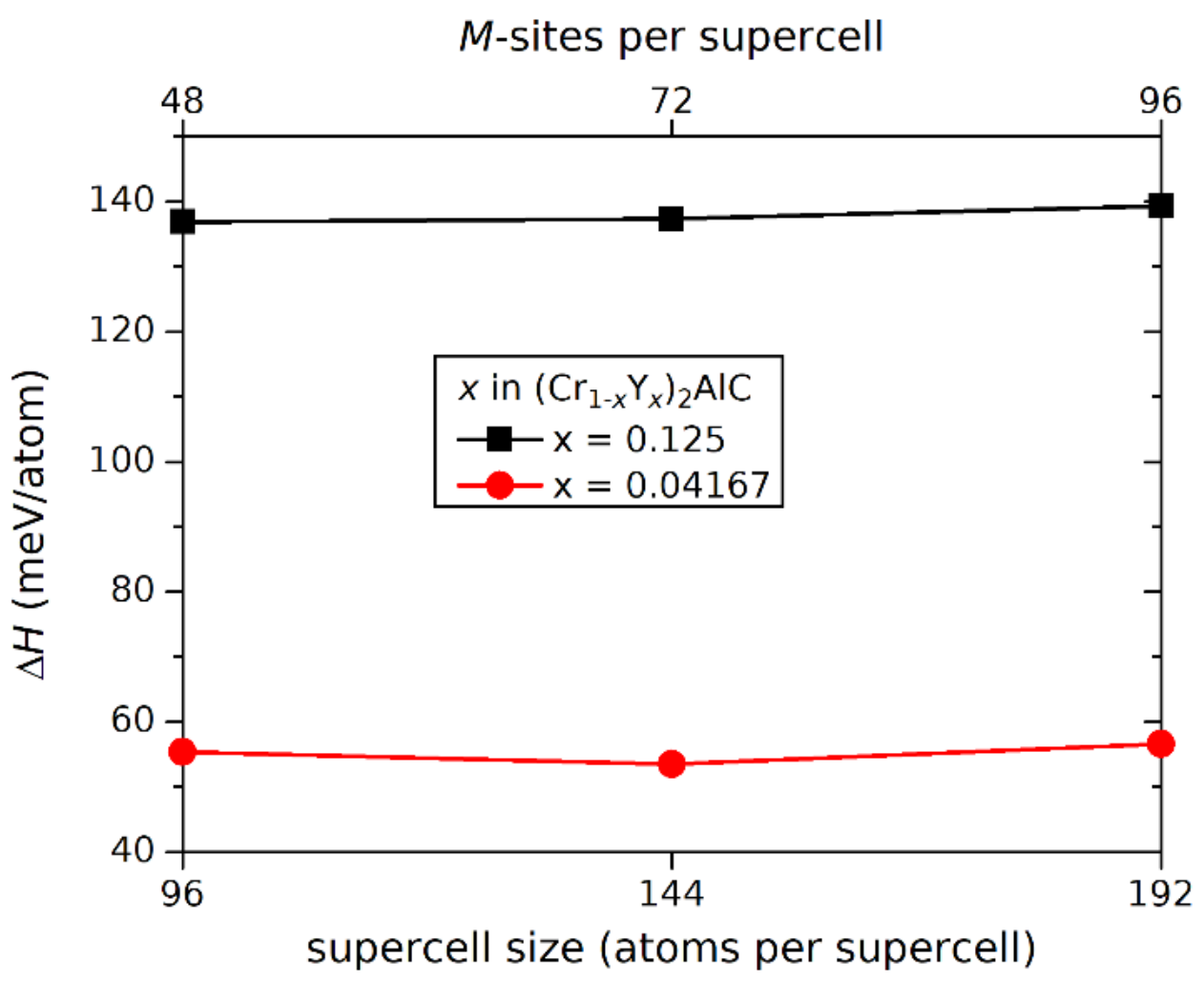

Figure S1. Calculated formation enthalpy for $(Cr_{1-x}Y_x)_2AlC$ as function of supercell size.

Table S1. Competing phases considered for the Cr-Y-Al-C quaternary system.

| Phase | Prototype | Space group |
|---|---|---|
| Y | Mg | $P6_3/mmc$ (194) |
| Cr | W | Im-3m (229) |
| Al | Cu | Fm-3m (225) |
| C | C (graphite) | $P6_3/mmc$ (194) |
| $Al_4C_3$ | $Al_4C_3$ | R-3m (166) |
| $Cr_{23}C_6$ | $Cr_{23}C_6$ | Fm-3m (225) |
| $Cr_3C$ | $Fe_3C$ | Pnma (62) |
| $Cr_7C_3$ | $Cr_7C_3$ | Pnma (62) |
| $Cr_3C_2$ | $Sb_2S_3$ | Pnma (62) |
| $Cr_2Al$ | $MoSi_2$ | I4/mmm (13) |
| $Cr_5Al_8$ | $Cu_5Zn_8$ | I-43m (217) |
| $Cr_4Al_9$ | PdAl | R3m (160) |
| $Cr_5Al_{21}$ | | R3m (160) |
| $Cr_7Al_{45}$ | $V_7Al_{45}$ | C2/m (12) |
| $Y_2C$ | $CdCl_2$ | R-3m h (166) |
| $Y_4C_5$ | $Y_4C_5$ | Pbam (55) |
| $Y_3C_4$ | $Sc_3C_4$ | P4/mnc (128) |
| $YC_2$ | $CaC_2$ | I4/mmm (139) |
| $Y_2Al$ | $Co_2Si$-b | Pnma (62) |
| $Y_3Al_2$ | $Zr_3Al_2$ | $P4_2/mnm$ (136) |
| YAl | TlI | Cmcm (63) |
| $YAl_2$ | $Cu_2Mg$ | Fd-3m (227) |
| $YAl_3$ | $Mg_3Cd$ | $P6_3$/mmc (194) |
| $Cr_3AlC$ | $CaTiO_3$ | Pm-3m (221) |
| $Cr_2AlC$ | $Cr_2AlC$ | $P6_3$/mmc (194) |
| $Cr_3AlC_2$ | $Ti_3SiC_2$ | $P6_3$/mmc (194) |
| $Cr_4AlC_3$ | $Ti_4AlN_3$ | $P6_3$/mmc (194) |
| $YAl_3C_3$ | $ScAl_3C_3$ | $P6_3$/mc (186) |
| $Y_3AlC$ | $CaTiO_3$ | Pm-3m (221) |
| $Y_2AlC$ | $Cr_2AlC$ | $P6_3$/mmc (194) |
| $Y_3AlC_2$ | $Ti_3SiC_2$ | $P6_3$/mmc (194) |
| $Y_4AlC_3$ | $Ti_4AlN_3$ | $P6_3$/mmc (194) |
| $YCr_4Al_8$ | $CeMn_4Al_8$ | I4/mmm (139) |
| $Y_6Cr_4Al_{43}$ | $Ho_6Mo_4Al_{43}$ | $P6_3$/mcm (193) |
| $Y_2Cr_2C_3$ | $Ho_2Cr_2C_3$ | C12/m1 (12) |

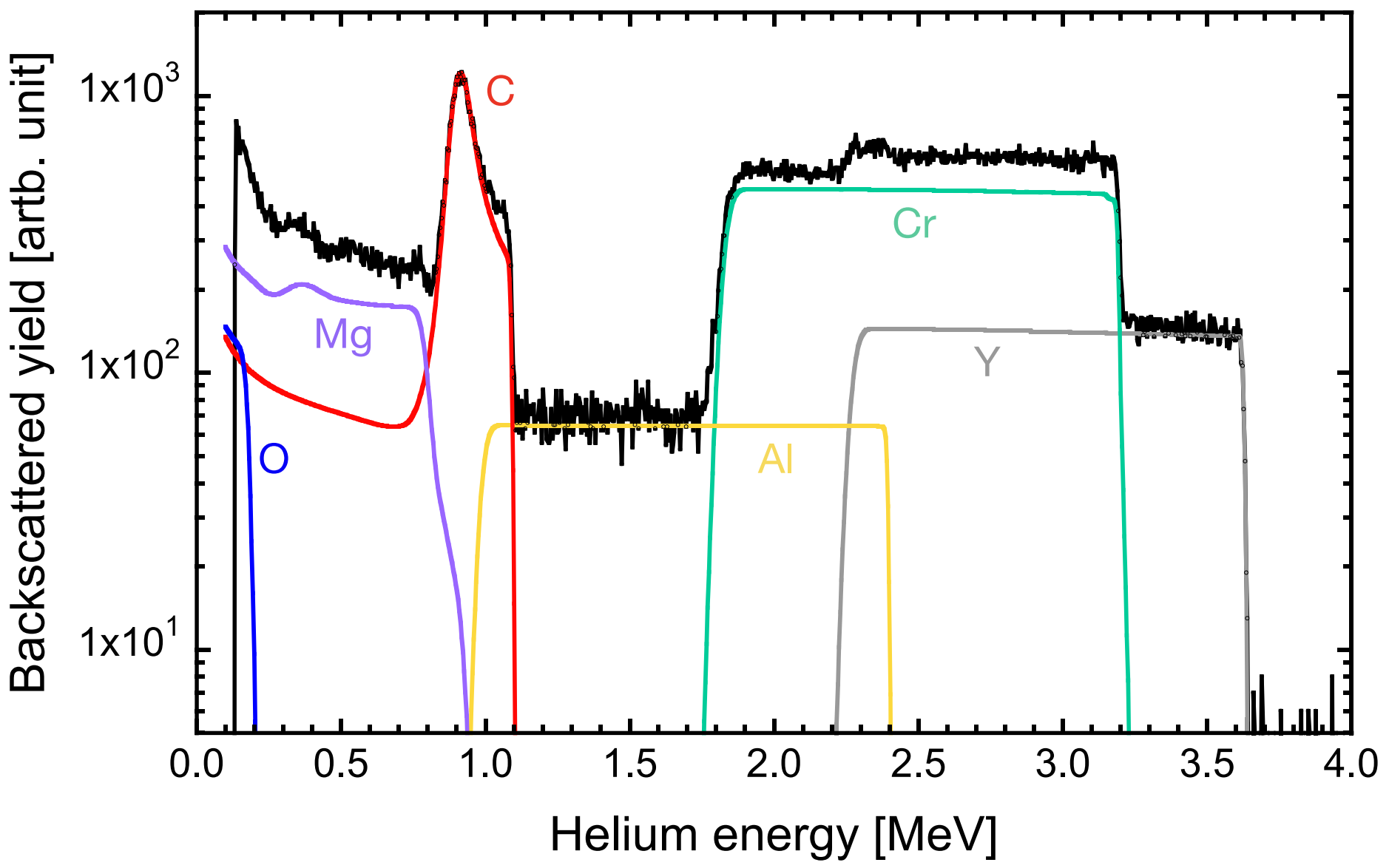

Figure S2: EBS spectrum of as deposited $Cr_2AlC$ with 5.4 at.% Y

## References

[1] M.W. Barsoum, T. El-Raghy, The MAX phases: Unique new carbide and nitride materials, Am. Sci. 89 (2001) 334–343. https://doi.org/10.1511/2001.4.334.

[2] P. Eklund, M. Beckers, U. Jansson, H. Högberg, L. Hultman, The Mn + 1AXn phases: Materials science and thin-film processing, Thin Solid Films. 518 (2010) 1851–1878. https://doi.org/10.1016/j.tsf.2009.07.184.

[3] M.W. Barsoum, M. Radovic, Elastic and Mechanical Properties of the MAX Phases, Annu. Rev. Mater. Res. (2011). https://doi.org/10.1146/annurev-matsci-062910-100448.

[4] C. Azina, S. Mráz, G. Greczynski, M. Hans, D. Primetzhofer, J.M. Schneider, P. Eklund, Oxidation behaviour of V2AlC MAX phase coatings, J. Eur. Ceram. Soc. 40 (2020) 4436–4444. https://doi.org/10.1016/j.jeurceramsoc.2020.05.080.

[5] S. Badie, D. Sebold, R. Vaßen, O. Guillon, J. Gonzalez-Julian, Mechanism for breakaway oxidation of the Ti2AlC MAX phase, Acta Mater. 215 (2021). https://doi.org/10.1016/j.actamat.2021.117025.

[6] J. Gonzalez-Julian, Processing of MAX phases: From synthesis to applications, J. Am. Ceram. Soc. 104 (2021) 659–690. https://doi.org/10.1111/jace.17544.
[7] D.E. Hajas, M. To Baben, B. Hallstedt, R. Iskandar, J. Mayer, J.M. Schneider, Oxidation of Cr2AlC coatings in the temperature range of 1230 to 1410°C, Surf. Coatings Technol. 206 (2011) 591–598. https://doi.org/10.1016/j.surfcoat.2011.03.086.
[8] B. Cui, W.E. Lee, B. Cui, W.E. Lee, High-temperature Oxidation Behaviour of MAX Phase Ceramics, Refract. Worldforum. 5 (2013) 105–112.
[9] E. Charalampopoulou, K. Lambrinou, T. Van der Donck, B. Paladino, F. Di Fonzo, C. Azina, P. Eklund, S. Mráz, J.M. Schneider, D. Schryvers, R. Delville, Early stages of dissolution corrosion in 316 L and DIN 1.4970 austenitic stainless steels with and without anticorrosion coatings in static liquid lead-bismuth eutectic (LBE) at 500 °C, Mater. Charact. 178 (2021) 111234. https://doi.org/10.1016/j.matchar.2021.111234.
[10] B. Anasori, M.R. Lukatskaya, Y. Gogotsi, 2D metal carbides and nitrides (MXenes) for energy storage, 2017. https://doi.org/10.1038/natrevmats.2016.98.
[11] B. Anasori, Y. Gogotsi, 2D Metal carbides and nitrides (MXenes): Structure, properties and applications, Springer International Publishing, 2019. https://doi.org/10.1007/978-3-030-19026-2.
[12] B. Tunca, T. Lapauw, O.M. Karakulina, M. Batuk, T. Cabioc'h, J. Hadermann, R. Delville, K. Lambrinou, J. Vleugels, Synthesis of MAX Phases in the Zr-Ti-Al-C System, Inorg. Chem. 56 (2017) 3489–3498. https://doi.org/10.1021/acs.inorgchem.6b03057.
[13] T. Lapauw, B. Tunca, D. Potashnikov, A. Pesach, O. Ozeri, J. Vleugels, K. Lambrinou, The double solid solution (Zr, Nb)2(Al, Sn)C MAX phase: a steric stability approach, Sci. Rep. 8 (2018) 1–13. https://doi.org/10.1038/s41598-018-31271-2.
[14] J. Lu, A. Thore, R. Meshkian, Q. Tao, L. Hultman, J. Rosen, Theoretical and Experimental Exploration of a Novel In-Plane Chemically Ordered (Cr2/3M1/3)2AlC i-MAX Phase with M = Sc and y, Cryst. Growth Des. 17 (2017) 5704–5711. https://doi.org/10.1021/acs.cgd.7b00642.
[15] C. Azina, P. Eklund, Effects of temperature and target power on the sputter-deposition of (Ti,Zr)n+1AlCn MAX-phase thin films, Results Mater. 9 (2021) 100159. https://doi.org/10.1016/j.rinma.2020.100159.
[16] C. Azina, B. Tunca, A. Petruhins, B. Xin, M. Yildizhan, P.O.Å. Persson, J. Vleugels, K. Lambrinou, J. Rosen, P. Eklund, Deposition of MAX phase-containing thin films from a (Ti,Zr)2AlC compound target, Appl. Surf. Sci. 551 (2021) 149370. https://doi.org/10.1016/j.apsusc.2021.149370.
[17] T.A. ElMeligy, E. Epifano, M. Sokol, G. Hug, M. Hans, J.M. Schneider, M.W. Barsoum, Isothermal Oxidation of Ti3Al0.6Ga0.4C2 MAX Phase Solid Solution in Air at 1000 °C to 1300 °C, J. Electrochem. Soc. 169 (2022) 031510. https://doi.org/10.1149/1945-7111/AC58C1.
[18] K.G. Pradeep, K. Chang, A. Kovács, S. Sen, A. Marshal, R. de Kloe, R.E. Dunin-Borkowski, J.M. Schneider, Nano-scale Si segregation and precipitation in Cr 2 Al(Si)C MAX phase coatings impeding grain growth during oxidation, Mater. Res. Lett. 7 (2019) 180–187. https://doi.org/10.1080/21663831.2019.1572663.
[19] T.A. ElMelegy, M. Sokol, M.W. Barsoum, Enhanced yield synthesis of bulk

dense (M2/3Y1/3)2AlC (M = Cr, W, Mo) in-plane chemically ordered quaternary atomically laminated i-MAX phases and oxidation of (Cr2/3Y1/3)2AlC and (Mo2/3Y1/3)2AlC, J. Alloys Compd. 867 (2021) 158930. https://doi.org/10.1016/j.jallcom.2021.158930.
[20] O. Berger, C. Leyens, S. Heinze, R. Boucher, M. Ruhnow, Characterization of Cr-Al-C and Cr-Al-C-Y films synthesized by High Power Impulse Magnetron Sputtering at a low deposition temperature, Thin Solid Films. 580 (2015) 6–11. https://doi.org/10.1016/j.tsf.2015.03.008.
[21] O. Berger, R. Boucher, M. Ruhnow, Part II. Oxidation of yttrium doped Cr2AlC films in temperature range between 700 and 1200°C, Surf. Eng. 31 (2015) 386–396. https://doi.org/10.1179/1743294414Y.0000000418.
[22] G. Kresse, J. Furthmüller, Efficient iterative schemes for *ab initio* total-energy calculations using a plane-wave basis set, Phys. Rev. B. 54 (1996) 11169. https://doi.org/10.1103/PhysRevB.54.11169.
[23] J.P. Perdew, K. Burke, M. Ernzerhof, Generalized Gradient Approximation Made Simple, Phys. Rev. Lett. 77 (1996) 3865. https://doi.org/10.1103/PhysRevLett.77.3865.
[24] H.J. Monkhorst, J.D. Pack, Special points for Brillonin-zone integrations*, NUMBER. 13 (1976).
[25] A. Zunger, S.H. Wei, L.G. Ferreira, J.E. Bernard, Special quasirandom structures, Phys. Rev. Lett. 65 (1990) 353. https://doi.org/10.1103/PhysRevLett.65.353.
[26] A. Van De Walle, P. Tiwary, M. De Jong, D.L. Olmsted, M. Asta, A. Dick, D. Shin, Y. Wang, L.Q. Chen, Z.K. Liu, Efficient stochastic generation of special quasirandom structures, Calphad. 42 (2013) 13–18. https://doi.org/10.1016/J.CALPHAD.2013.06.006.
[27] M. Dahlqvist, B. Alling, I.A. Abrikosov, J. Rosén, Phase stability of Ti 2 AlC upon oxygen incorporation: A first-principles investigation, (n.d.). https://doi.org/10.1103/PhysRevB.81.024111.
[28] M. Dahlqvist, B. Alling, J. Rosén, Stability trends of MAX phases from first principles, Phys. Rev. B. 81 (n.d.). https://doi.org/10.1103/PhysRevB.81.220102.
[29] A. Thore, M. Dahlqvist, B. Alling, J. Rosén, Temperature dependent phase stability of nanolaminated ternaries from first-principles calculations, Comput. Mater. Sci. 91 (2014) 251–257. https://doi.org/10.1016/J.COMMATSCI.2014.04.055.
[30] R. Meshkian, Q. Tao, M. Dahlqvist, J. Lu, L. Hultman, J. Rosen, Theoretical stability and materials synthesis of a chemically ordered MAX phase, Mo2ScAlC2, and its two-dimensional derivate Mo2ScC2 MXene, Acta Mater. 125 (2017) 476–480. https://doi.org/10.1016/J.ACTAMAT.2016.12.008.
[31] M. Dahlqvist, J. Lu, R. Meshkian, Q. Tao, L. Hultman, J. Rosen, Prediction and synthesis of a family of atomic laminate phases with Kagomé-like and in-plane chemical ordering, Sci. Adv. 3 (2017). https://doi.org/10.1126/SCIADV.1700642/SUPPL_FILE/1700642_SM.PDF.
[32] B. Anasori, M. Dahlqvist, J. Halim, E.J. Moon, J. Lu, B.C. Hosler, E.N. Caspi, S.J. May, L. Hultman, P. Eklund, J. Rosén, M.W. Barsoum, Experimental and theoretical characterization of ordered MAX phases Mo2TiAlC2 and Mo2Ti2AlC3, J. Appl. Phys. 118 (2015) 094304. https://doi.org/10.1063/1.4929640.

[33] A.S. Ingason, A. Petruhins, M. Dahlqvist, F. Magnus, A. Mockute, B. Alling, L. Hultman, I.A. Abrikosov, P.O.Å. Persson, J. Rosen, A nanolaminated magnetic phase: Mn2GaC, Mater. Res. Lett. 2 (2017) 89–93. https://doi.org/10.1080/21663831.2013.865105/SUPPL_FILE/TMRL_A_865105_SM4991.PDF.

[34] P. Eklund, M. Dahlqvist, O. Tengstrand, L. Hultman, J. Lu, N. Nedfors, U. Jansson, J. Rosén, Discovery of the ternary nanolaminated compound Nb2GeC by a systematic theoretical-experimental approach, Phys. Rev. Lett. 109 (2012). https://doi.org/10.1103/PHYSREVLETT.109.035502.

[35] M. Dahlqvist, B. Alling, I.A. Abrikosov, J. Rosen, Magnetic nanoscale laminates with tunable exchange coupling from first principles, Phys. Rev. B - Condens. Matter Mater. Phys. 84 (2011) 220403. https://doi.org/10.1103/PHYSREVB.84.220403/FIGURES/4/MEDIUM.

[36] A.S. Ingason, A. Mockute, M. Dahlqvist, F. Magnus, S. Olafsson, U.B. Arnalds, B. Alling, I.A. Abrikosov, B. Hjörvarsson, P.O.. Persson, J. Rosen, Magnetic Self-Organized Atomic Laminate from First Principles and Thin Film Synthesis, Phys. Rev. Lett. 110 (2013) 195502. https://doi.org/10.1103/PhysRevLett.110.195502.

[37] P. Hohenberg, W. Kohn, Inhomogeneous electron gas, Phys. Rev. 136 (1964) B864. https://doi.org/10.1103/PHYSREV.136.B864/FIGURE/1/THUMB.

[38] W. Kohn, L.J. Sham, Self-consistent equations including exchange and correlation effects, Phys. Rev. 140 (1965) A1133. https://doi.org/10.1103/PHYSREV.140.A1133/FIGURE/1/THUMB.

[39] G. Kresse, J. Furthmüller, Efficiency of ab-initio total energy calculations for metals and semiconductors using a plane-wave basis set, Comput. Mater. Sci. 6 (1996) 15–50. https://doi.org/10.1016/0927-0256(96)00008-0.

[40] M. Dahlqvist, B. Alling, J. Rosén, Correlation between magnetic state and bulk modulus of Cr2AlC, J. Appl. Phys. 113 (2013) 216103. https://doi.org/10.1063/1.4808239.

[41] S.L. Dudarev, G.A. Botton, S.Y. Savrasov, C.J. Humphreys, A.P. Sutton, Electron-energy-loss spectra and the structural stability of nickel oxide: An LSDA+U study, Phys. Rev. B. 57 (1998) 1505. https://doi.org/10.1103/PhysRevB.57.1505.

[42] Genhringer, sqsgenerator 0.2 documentation, (n.d.). https://sqsgenerator.readthedocs.io/en/latest/index.html (accessed July 4, 2022).

[43] M. Samuelsson, D. Lundin, J. Jensen, M.A. Raadu, J.T. Gudmundsson, U. Helmersson, On the film density using high power impulse magnetron sputtering, Surf. Coatings Technol. 205 (2010) 591–596. https://doi.org/10.1016/J.SURFCOAT.2010.07.041.

[44] G. Greczynski, J. Lu, S. Bolz, W. Kölker, C. Schiffers, O. Lemmer, I. Petrov, J.E. Greene, L. Hultman, Novel strategy for low-temperature, high-rate growth of dense, hard, and stress-free refractory ceramic thin films, J. Vac. Sci. Technol. A Vacuum, Surfaces, Film. 32 (2014) 041515. https://doi.org/10.1116/1.4884575.

[45] V. Schnabel, M. Köhler, S. Evertz, J. Gamcova, J. Bednarcik, D. Music, D. Raabe, J.M. Schneider, Revealing the relationships between chemistry, topology and stiffness of ultrastrong Co-based metallic glass thin films: A combinatorial approach, Acta Mater. 107 (2016) 213–219. https://doi.org/10.1016/j.actamat.2016.01.060.

[46] P. Ström, D. Primetzhofer, Ion beam tools for nondestructive in-situ and in-operando composition analysis and modification of materials at the Tandem Laboratory in Uppsala, J. Instrum. 17 (2022) P04011. https://doi.org/10.1088/1748-0221/17/04/P04011.

[47] B. Stelzer, X. Chen, P. Bliem, M. Hans, B. Völker, R. Sahu, C. Scheu, D. Primetzhofer, J.M. Schneider, Remote Tracking of Phase Changes in Cr2AlC Thin Films by In-situ Resistivity Measurements, Sci. Reports 2019 91. 9 (2019) 1–7. https://doi.org/10.1038/s41598-019-44692-4.

[48] M.S. Janson, CONTES instruction manual, Uppsala, 2004.

[49] J.A. Leavitt, L.C. McIntyre, M.D. Ashbaugh, J.G. Oder, Z. Lin, B. Dezfouly-Arjomandy, Cross sections for 170.5° backscattering of 4He from oxygen for 4He energies between 1.8 and 5.0 MeV, Nucl. Instruments Methods Phys. Res. Sect. B Beam Interact. with Mater. Atoms. 44 (1990) 260–265. https://doi.org/10.1016/0168-583X(90)90637-A.

[50] M. Mayer, SIMNRA, a simulation program for the analysis of NRA, RBS and ERDA, AIP Conf. Proc. 475 (2008) 541. https://doi.org/10.1063/1.59188.

[51] A.C. Dippel, H.P. Liermann, J.T. Delitz, P. Walter, H. Schulte-Schrepping, O.H. Seeck, H. Franz, Beamline P02.1 at PETRA III for high-resolution and high-energy powder diffraction, J. Synchrotron Radiat. 22 (2015) 675–687. https://doi.org/10.1107/S1600577515002222/CO5063SUP1.PDF.

[52] A. Mockute, M. Dahlqvist, J. Emmerlich, L. Hultman, J.M. Schneider, P.O.˚ A. Persson, J. Rosen, Synthesis and ab initio calculations of nanolaminated (Cr,Mn) 2 AlC compounds, Phys. Rev. B. 87 (2013) 94113. https://doi.org/10.1103/PhysRevB.87.094113.

[53] H. Pazniak, M. Stevens, M. Dahlqvist, B. Zingsem, L. Kibkalo, M. Felek, S. Varnakov, M. Farle, J. Rosen, U. Wiedwald, Phase Stability of Nanolaminated Epitaxial (Cr1- xFex)2AlC MAX Phase Thin Films on MgO(111) and Al2O3(0001) for Use as Conductive Coatings, ACS Appl. Nano Mater. 4 (2021) 13761–13770. https://doi.org/10.1021/ACSANM.1C03166/ASSET/IMAGES/LARGE/AN1C03166_0008.JPEG.

[54] A. Abdulkadhim, M. to Baben, T. Takahashi, V. Schnabel, M. Hans, C. Polzer, P. Polcik, J.M. Schneider, Crystallization kinetics of amorphous Cr2AlC thin films, Surf. Coatings Technol. 206 (2011) 599–603. https://doi.org/10.1016/j.surfcoat.2011.06.003.

[55] R. Richter, Z. Altounian, J.O. Strom-Olsen, Y5Al3; a new Y-Al compound, J. Mater. Sci. 22 (1987) 2986.

[56] E. Drouelle, V. Gauthier-Brunet, J. Cormier, P. Villechaise, P. Sallot, F. Naimi, F. Bernard, S. Dubois, Microstructure-oxidation resistance relationship in Ti3AlC2 MAX phase, J. Alloys Compd. 826 (2020) 154062. https://doi.org/10.1016/j.jallcom.2020.154062.

[57] X. Chen, B. Stelzer, M. Hans, R. Iskandar, J. Mayer, J.M. Schneider, Enhancing the high temperature oxidation behavior of Cr2AlC coatings by reducing grain boundary nanoporosity, Mater. Res. Lett. 9 (2021) 127–133. https://doi.org/10.1080/21663831.2020.1854358.

[58] H. Rueß, M. to Baben, S. Mráz, L. Shang, P. Polcik, S. Kolozsvári, M. Hans, D. Primetzhofer, J.M. Schneider, HPPMS deposition from composite targets: Effect of two orders of magnitude target power density changes on the composition of sputtered Cr-Al-C thin films, Vacuum. 145 (2017) 285–289. https://doi.org/10.1016/J.VACUUM.2017.08.048.

[59] Y.-P. Chien, S. Mráz, M. Fekete, M. Hans, D. Primetzhofer, S. Kolozsvári, P. Polcik, J.M. Schneider, Deviations between film and target compositions induced by backscattered Ar during sputtering from M2-Al-C (M = Cr, Zr, and Hf) composite targets, Surf. Coatings Technol. 446 (2022) 128764. https://doi.org/10.1016/J.SURFCOAT.2022.128764.